\documentclass[prl,aps,twocolumn,raggedbottom,showpacs,nobalancelastpage,amssymb,groupeaddress]{revtex4}

\usepackage[english]{babel}
\usepackage{graphicx}
\usepackage{amssymb}
\usepackage{amsmath}
\usepackage{amscd}
\usepackage{eucal}
\usepackage{color}
\usepackage{bm}

\usepackage{natbib}
\usepackage[bookmarks=true,colorlinks,linkcolor=red,urlcolor=blue,citecolor=blue]{hyperref}
\usepackage{dcolumn}

\def\be{\begin{equation}}
\def\ee{\end{equation}}
\def\bea{\begin{eqnarray}}
\def\eea{\end{eqnarray}}
\def\bsub{\begin{subequations}}
\def\esub{\end{subequations}}

\usepackage{ulem}

\begin{document}

\title{Andreev magneto-interferometry in topological hybrid junctions}

\newcommand{\spsms}{CEA-INAC/UJF Grenoble 1, SPSMS UMR-E 9001, Grenoble F-38054, France}

\author{Pierre Carmier}
\affiliation{\spsms}
\date{\today}

\begin{abstract}
We investigate the influence of the superconducting (S) proximity effect in the quantum Hall (QH) regime by computing the charge conductance flowing through a graphene-based QH/S/QH junction. This situation offers the exciting possibility of studying the fate of topological edge states when they experience tunneling processes through the superconductor. We predict the appearance of conductance peaks at integer values of the Landau level filling factor, as a consequence of the quantum interferences taking place at the junction, and provide a semiclassical analysis allowing for a natural interpretation of these interferences in terms of electron and hole trajectories propagating along the QH/S interfaces. Our results suggest that non-trivial junctions between topologically distinct phases could offer a highly tunable means of partitioning the flow of edge states.
\end{abstract}


\maketitle

Quantum Hall (QH) and superconducting (S) proximity effects are two prominent mesoscopic phenomena, yet their interplay has received little attention so far due to the widespread assumption that their respective ranges of validity are incompatible. However, in modern high mobility two-dimensional electron gases (with mean free paths typically exceeding the micron scale), magnetic field values $B$ required to enter the QH regime have become sufficiently small to allow the fabrication of QH/S junctions, using high critical field superconductors such as Nb compounds \cite{Eroms05}. These junctions feature chiral edge states of mixed electron-hole nature due to the Andreev reflection experienced by carriers at the interface with the S region \cite{Hoppe00}. Evidence for such edge states was successfully demonstrated a few years ago in InAs semiconducting heterostructures \cite{Eroms05}, following seminal experiments \cite{Takayanagi98, Moore99} and earlier theoretical proposals \cite{Takagaki98,Asano00,Chtchelkatchev01,Hoppe00}. The advent of graphene, in which both QH and S proximity effects have been routinely observed \cite{Novoselov05,Zhang05,Heersche07} owing to the material's low cost and tunability, has led to a revival of experimental activity in the field very recently \cite{Komatsu12,Popinciuc12,Rickhaus12}. 

\begin{figure}[]
\begin{center}
\includegraphics[angle=0,width=1.0\linewidth]{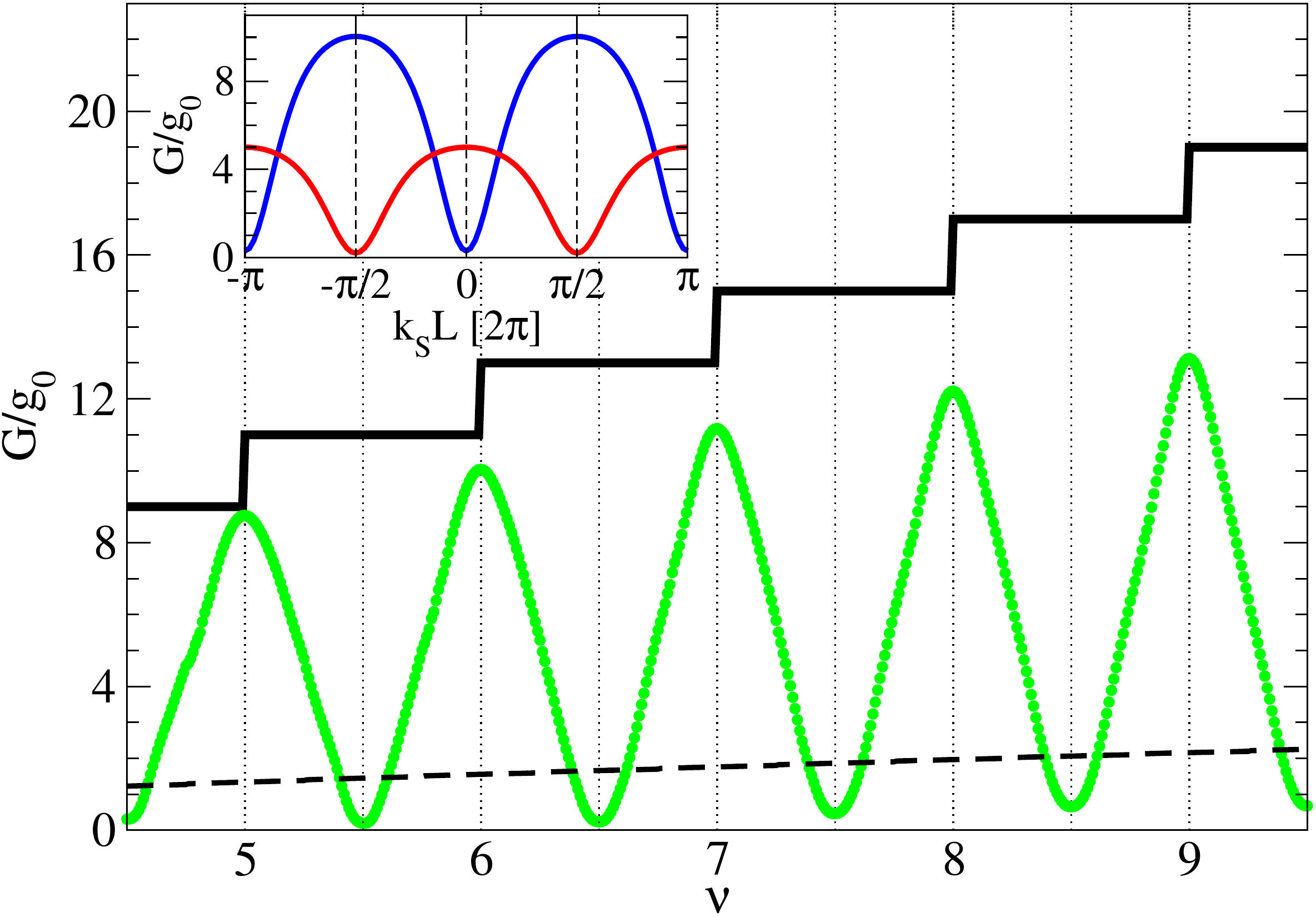}
\caption{(Color online): Conductance $G$ flowing through a QH/S/QH junction of length $L/\xi_S=2$ and width $W/l_B=40$ as a function of the Landau level filling factor $\nu$ for $k_SL=\pi/2 \; [2\pi]$ (green circles). Transmission peaks occur at integer values of $\nu$. Regular QH plateaus (when the S region turns normal) and classical expectation are plotted for comparison (thick black and dashed black lines, respectively). Inset: Value of the peak at $\nu=6$ (top blue) and of the dip at $\nu=5.5$ (bottom red) as a function of $k_SL$.}
\label{FigInterfero}
\end{center}
\end{figure}

The interface between a QH insulator and a superconductor actually provides a particularly interesting and non-trivial realization of a topological junction, which is a junction between bulk insulating phases (from a single-particle perspective) characterized by different topological invariants \cite{Hasan10,Qi11}. A defining property of these junctions is the existence of topologically protected edge states propagating along their interface. Partition of the information carried by these states in the available outgoing channels of a mesoscopic system is an important problem, both from a fundamental point of view and in the perspective of exploiting the robustness of topological phases to build novel electronic devices. The purpose of this Letter is to highlight the potentially crucial role played by quantum interferences in such topological junctions, a spectacular example of which is depicted in Fig.~\ref{FigInterfero} where the charge conductance flowing through a QH/S/QH junction is plotted as a function of the Landau level filling factor $\nu$. 
Instead of featuring plateaus at odd values of the spin-degenerate conductance quantum $g_0=2e^2/h$ (thick black line), as would be the case in the absence of the S region (or equivalently for supercritical magnetic fields), the conductance is seen to oscillate between extremal values, from a situation where current is essentially blocked to one where it is fully transmitted through the S region. In the following, I will provide a quantitative understanding of this phenomenon, using the classical picture of skipping orbits to describe QH edge states in the ballistic regime \cite{Chtchelkatchev01,Chtchelkatchev07,Carmier10,Carmier11}. By expressing the conductance as a sum over the various semiclassical trajectories contributing to the transmission probability through the QH/S/QH junction, we will see that the magneto-oscillations featured in Fig.~\ref{FigInterfero} can be naturally interpreted in terms of interferences between electron and hole paths propagating along the QH/S interfaces. While suspended graphene should be a well-suited candidate to test these predictions, as demonstrated by recent experimental evidence supporting phase-coherent ballistic transport in this system \cite{Rickhaus13,Grushina13,Mizuno13}, the obtained results are essentially independent of graphene's band structure and should therefore be observable in other two-dimensional electron gases as well.

\begin{figure}[]
\begin{center}
\includegraphics[angle=0,width=1.0\linewidth]{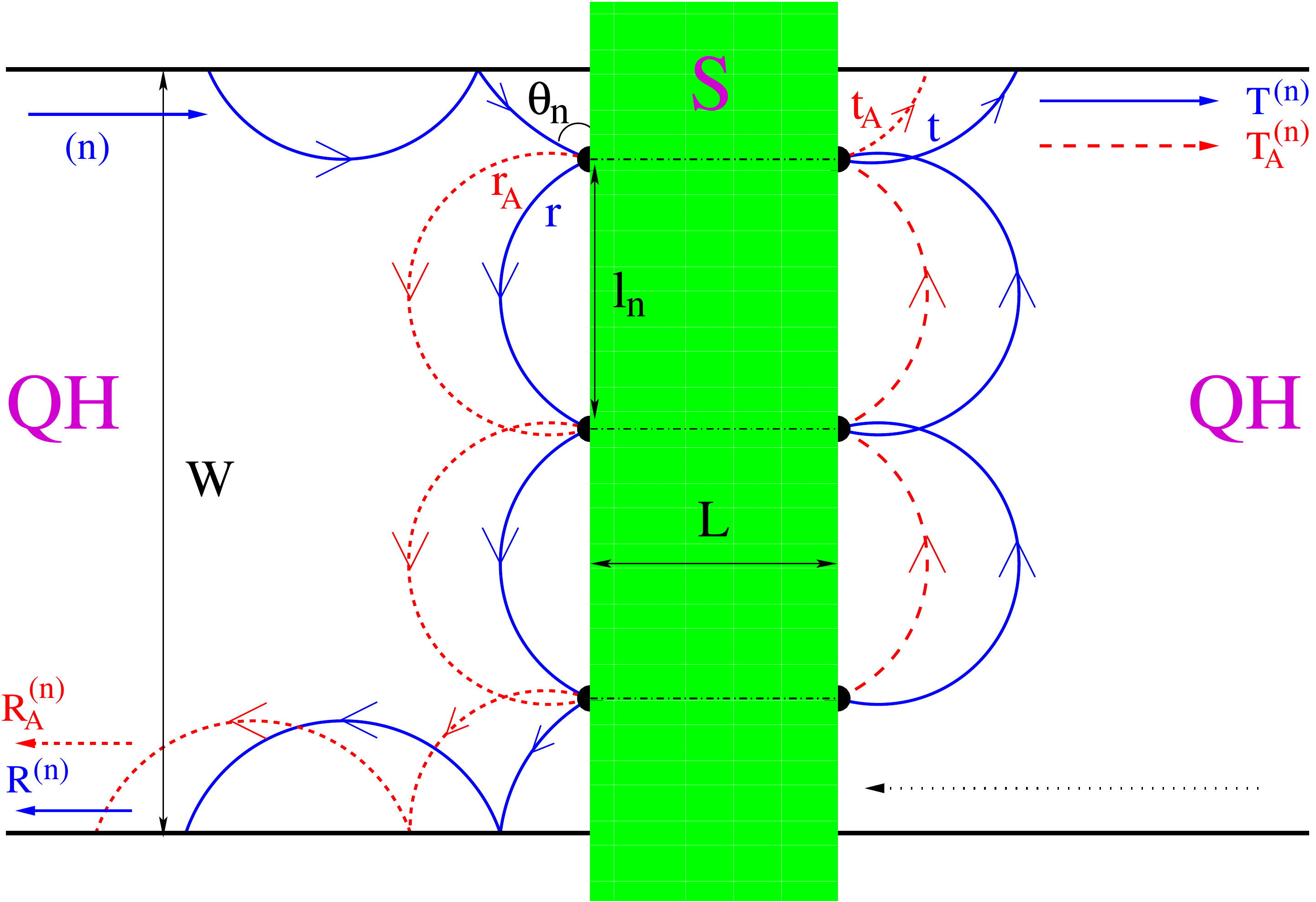}
\caption{(Color online): Cartoon of a QH/S/QH junction with $N=3$ vertices. Full blue lines are for electron trajectories and dashed red lines for hole trajectories. Semiclassically, states along the QH/S interfaces can be described by skipping orbits propagating between equidistant vertices. Given an incoming mode $(n)$ on the upper left edge, we seek how the outgoing probabilities $R^{(n)}$, $R_A^{(n)}$, $T^{(n)}$, $T_A^{(n)}$ scale with $W$.}
\label{FigSetup}
\end{center}
\end{figure}

Let us consider the geometry depicted in Fig.~\ref{FigSetup}, where a spin-singlet superconductor (connected to a hidden S reservoir) is deposited on top of a two-terminal graphene ribbon of width $W$ in the QH regime, thereby opening a proximity-induced superconducting gap $\Delta_S$ (which will be assumed constant) in a strip of length $L$ inside the ribbon. 
Assuming phase-coherent ballistic transport inside the system, finite temperature $k_BT<k_BT_c$ (where $T_c\approx0.6\Delta_S/k_B$ is the critical temperature of the superconductor) should play no role beyond renormalizing the values of the superconducting gap and the critical field, and $k_BT$ will thus henceforth be set to zero. 
In (linear) response to a subgap bias voltage $eV<\Delta_S$ applied in the left lead, tunneling processes through the S region allow for a charge current to flow in the right lead, characterized by the electrical conductance \cite{Buettiker85,Blonder82}
\be
\label{eq:BTK}
G = g_0 \sum_n \left(T^{(n)}-T_A^{(n)}\right) \; ,
\ee
with $T^{(n)}$ the probability for mode $n$ to be transmitted as an electron and $T_A^{(n)}$ the probability to be transmitted as a hole (see Fig.~\ref{FigSetup}). Mode $n$ can also be reflected as an electron or as a hole with probabilities $R^{(n)}$ and $R_A^{(n)}$, such that $R^{(n)}+R_A^{(n)}+T^{(n)}+T_A^{(n)}=1$. 

In order to derive semiclassical approximations for these probabilities, one must describe both the incoming QH edge states and their dynamics at the interface with the S region semiclassically. The first part is rather straightforward. Provided $\nu=(k_Fl_B)^2 \gg 1$, where $k_F$ is the Fermi wavevector and $l_B=\sqrt{\hbar/(eB)}$ the magnetic length, the classical skipping orbit picture can be translated into a more rigorous semiclassical description by applying a Bohr-Sommerfeld quantization procedure to the periodic motion of the electrons, which effectively results in turning the continuous family of possible classical trajectories into a set of edge modes, each characterized by a quantized angle $0<\theta_{n,\pm}<\pi$. For an armchair edge, this quantization condition can be expressed as $2\theta_{n,\pm}-\sin{2\theta_{n,\pm}}=(2\pi/\nu)(n \pm 1/4)$ \cite{Carmier11,Rakyta10}, where $n$ can be identified with the Landau level index, and the $\pm$ sign refers to the lifting of the two-fold valley degeneracy in graphene (which will henceforth be implicit). However the choice of boundary condition at the edges of the ribbon is qualitatively unimportant for our purposes in the semiclassical regime $\nu\gg1$, where the low-energy selection rules imposed by the valley-polarization constraints characterizing the single channel case can be relaxed \cite{Akhmerov07}.

Let us now address the dynamics along the interface. Because the Lorentz force acting on a hole is the same as that acting on an electron, both particles rotate in the same direction with the magnetic field, leading to unidirectional motion along a given QH/S interface (see Fig.~\ref{FigSetup}). This is due to the fact that even though a hole carries an opposite charge from that of an electron, this is compensated by the hole's direction of motion being opposite to its momentum. The situation becomes more complicated in a QH/S/QH setup, as tunneling processes through the S region give rise to states localized along the second interface which counter-propagate with respect to those on the first interface (see Fig.~\ref{FigSetup}). The problem we face therefore boils down to describing the dynamics of coupled counter-propagating QH states (of mixed electron-hole nature). For simplicity, we shall restrict our analysis to the zero-bias limit, for which the cyclotron radii of electron and hole channels match. In this case, the semiclassical formalism is more tractable, since the ensemble of classical trajectories reduces to scattering between equidistant vertices along the interface (see Fig.~\ref{FigSetup}) -- assuming a sufficiently doped S region that the position mismatch between scattering vertices on both sides of the S region 
can be neglected. This approximation remains valid when a non-vanishing bias voltage $eV \lesssim \hbar v_F/W$ is taken into account, and the results can in principle be extended to arbitrary $eV$ \cite{Carmier13NSN} using the more technical Green's function approach based on the Fisher-Lee formula \cite{Fisher81,Baranger89,Carmier11}.

To proceed further, it is convenient to introduce the vector $(e_i,h_i)^T$ composed of electron and hole probability amplitudes of leaving vertex $i$ along the left side of the interface. Its evolution is governed by the 2 x 2 matrix ${\cal W}_i$, according to the equation $(e_{i+1}, h_{i+1})^T={\cal W}_i(e_i, h_i)^T$. All of the information regarding the semiclassical dynamics at vertex $i$ is encoded in ${\cal W}_i$ which will be referred to as the local propagator. Noting $N=[W/l_n]$ the (integer) number of vertices, where $l_n=2l_B\sqrt{2\nu}\sin{\theta_n}$ is the distance separating consecutive scattering events along the interface, it is clear that 
\be
\label{eq:Propag}
\left( \begin{array}{l} e_N \\ h_N \end{array} \right) = \prod_{i=1}^N {\cal W}_i \left( \begin{array}{l} 1 \\ 0 \end{array} \right) \; .
\ee
Transmission probabilities can then easily be obtained by summing over all possible coordinates for the initial scattering vertex: for example, 
\be
R^{(n)} = \frac{1}{l_n}\int_0^{l_n} dl |e_{N(l)}|^2 \; ,
\ee
with $N(l)=1+[(W-l)/l_n]$, such that $N\leq N(l)\leq N+1$.

The main task we are left with is to compute the local propagator ${\cal W}_i$. In order to do so, let us now look in more detail at the scattering processes taking place at the QH/S interfaces between consecutive vertices. For an incoming particle with angle $\theta_n$ at a given vertex, these processes are described by the 2 x 2 matrices
\be
\label{Scatt2}
{\cal R} = \left( \begin{array}{cc} re^{i\phi} & r_A'e^{i\phi} \\ r_Ae^{i\phi'} & r'e^{i\phi'} \end{array} \right) \; , \; {\cal T} = \left( \begin{array}{cc} te^{i\phi} & t_A'e^{i\phi} \\ t_Ae^{i\phi'} & t'e^{i\phi'} \end{array} \right) \; ,
\ee
where ${\cal R}$ corresponds to reflection along a given interface and ${\cal T}$ to transmission from one interface to the other. Phases $\phi$ and $\phi'$ account for the action, Maslov index and Berry phase \cite{Carmier10,Carmier11} respectively acquired by electron and hole channels during their propagation between scattering events on the interface, such that $\delta\phi=\phi-\phi'=2\pi\nu$. 
The scattering coefficients in Eq.~(\ref{Scatt2}) are the local probability amplitudes of normal reflection ($r$), Andreev reflection ($r_A$), elastic cotunneling ($t$) and crossed Andreev reflection ($t_A$). 
Assuming that the QH/S interfaces are abrupt on the scale of $l_B$, the existence of the magnetic field can be locally ignored and the scattering coefficients can be determined by matching the quantum mechanical wavefunctions 
at the interfaces and making use of momentum conservation arising from translational invariance in the transverse direction. In the limit $L\gtrsim\xi_S$, where $\xi_S=\hbar v_F/\Delta_S$ is the superconducting coherence length, one obtains
\be
\label{eq:coeffs}
\left\lbrace
\begin{array}{l}
r=-\cos{\theta_n}
\; , 
\\
r_A=-i\sin{\theta_n}
\; ,
\\
t=2\sin{\theta_n}(\cos{k_SL}\sin{\theta_n}+i\sin{k_SL})e^{-L/\xi_S}
\; ,
\\
t_A=-i\sin{2\theta_n}\cos{k_SL} \; e^{-L/\xi_S}
\; ,
\end{array}
\right.
\ee
assuming once more $k_S\gg k_F$, with $k_S$ the Fermi wavevector in the S region (primed coefficients in Eq.~(\ref{Scatt2}) can be obtained from those of Eq.~(\ref{eq:coeffs}) by reversing the sign of $k_S$).
These coefficients carry the signature of graphene's unusual band structure through their angular dependence \cite{Beenakker06}. In particular, $r$ and $t_A$ vanish under normal incidence as a consequence of the absence of backscattering (so-called Klein tunneling). Also, note that $t$ and $t_A$ are, as expected, exponentially suppressed on the scale of $\xi_S$. Therefore, in order for tunneling processes to play a role, typical lengths of the S region will be limited to sizes such that diamagnetic screening currents can be neglected \cite{Tinkham}. As a consequence, the validity of the plane wave approximation to tunneling coefficients in Eq.~(\ref{eq:coeffs}) will require magnetic lengths $l_B \gtrsim L$.

We now have all the necessary ingredients to compute the local propagator. Before presenting the general solution, let us familiarize ourselves with it by computing the first couple of terms. The first one, ${\cal W}_1={\cal R}$, is rather obvious: incoming carriers  at the first vertex must necessarily be reflected, else they will be transmitted through the S region and irrevocably leave the interface (see Fig.~\ref{FigSetup}). The second term can be expressed as an infinite sum, ${\cal W}_2 = {\cal R}\sum_{m=0}^{+\infty}{\cal T}^{2m}$, where the $m^{th}$ contribution takes into account trajectories where charge carriers have tunneled $2m$ times through the S region. It can be conveniently rewritten in a self-consistent form, ${\cal W}_2 = {\cal R} + {\cal W}_2{\cal T}^2$, the meaning of which is the following: unless carriers incoming at the vertex are directly reflected (${\cal R}$), they must tunnel twice through the S region (${\cal T}^2$), at which point one is back to the starting point (${\cal W}_2$). Elaborating on this idea, a general recurrence relation can be derived for $i\geq2$,
\be
\label{eq:GenRec}
{\cal W}_i = {\cal R} + \sum_{j=0}^{i-2}(\prod_{k=0}^j {\cal W}_{i-k}){\cal T}{\cal R}^j{\cal T} \; .
\ee
Likewise, an equation similar to Eq.~(\ref{eq:Propag}) can be written down for the electron and hole probability amplitudes to leave the QH/S/QH junction on the right side, $(e'_N, h'_N)^T={\cal W}'_N(1, 0)^T$, with ${\cal W}'_1 = {\cal T}$ and, for $N\geq2$,
\be
\label{eq:Rec2}
{\cal W}'_N =  {\cal T} + \sum_{j=0}^{N-2}(\prod_{k=0}^j {\cal W}_{N-k}){\cal T}{\cal R}^{j+1} \; .
\ee
In the limit $L/\xi_S \gg 1$ of a single QH/S interface,
Eq.~(\ref{eq:GenRec}) reduces to the uniform solution ${\cal W}_i={\cal R}$ describing the periodic skipping orbit motion along the left interface, and one thus retrieves the solution independently obtained by Chtchelkatchev in non-relativistic two-dimensional electron gases \cite{Chtchelkatchev01,Chtchelkatchev07} and by the author in graphene QH bipolar junctions \cite{Carmier10,Carmier11}.

\begin{figure}[]
\begin{center}
\includegraphics[angle=0,width=1.0\linewidth]{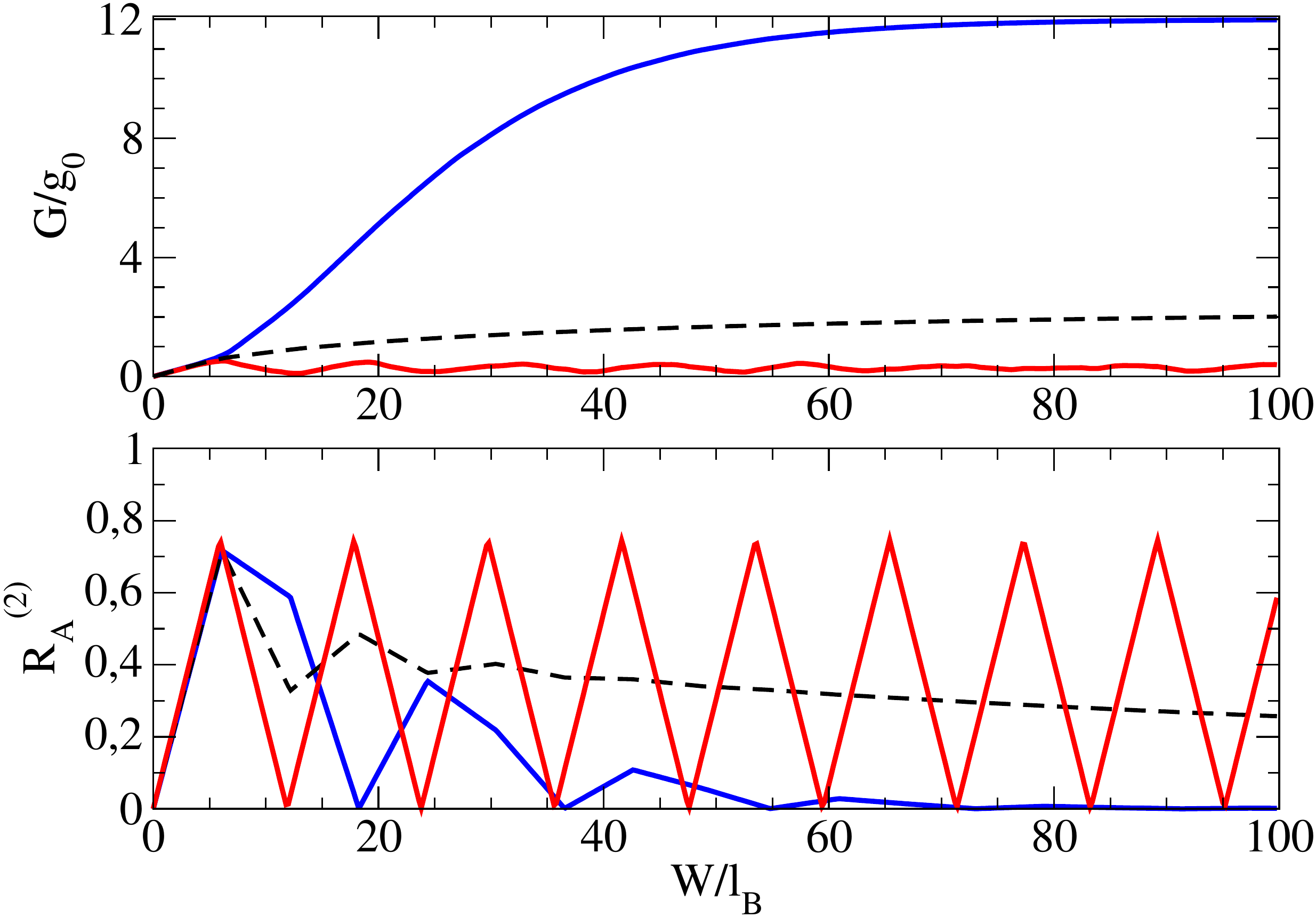}
\caption{(Color online): Upper panel: $G$ as a function of $W$ for $L/\xi_S=2$, $k_SL=\pi/2 \; [2\pi]$, and $\nu=6$ (top blue), $\nu=5.5$ (bottom red). Lower panel: Likewise for the probability of a given mode ($n=2$) to be reflected as a hole for $\nu=6$ (bottom blue), $\nu=5.5$ (top red). Dashed black lines are the classical expectations.
}
\label{FigRetro}
\end{center}
\end{figure}

Eqs.~(\ref{eq:GenRec}, \ref{eq:Rec2}) are the central results of this Letter and can be solved numerically. There are special cases, however, where they can be exactly solved, an important one for our purposes being when the condition $t+t'e^{-i\delta\phi}=0$ is fulfilled, which is equivalent to
\be
\label{eq:rescond}
\tan{\pi\nu}=\frac{\sin{\theta_n}}{\tan{k_SL}} \; .
\ee
In this case, the local propagator can be shown to take the simple form ${\cal W}_i=\alpha_i{\cal R}$, with $0<\alpha_i\leq 1$, and it is then a simple task to show that this implies $R^{(n)}+R_A^{(n)} \leq e^{-N(|t|^2+|t_A|^2)}$: in other words, full transmission of current through the S region is achieved exponentially fast with $W$ (blue curves in Fig.~\ref{FigRetro}). 
One can also easily prove that ${\cal W}'_N=\alpha'_N{\cal T}$ if Eq.~(\ref{eq:rescond}) holds, thereby yielding for the asymptotic value of transmission in the S region
\be
\label{eq:asympt}
T^{(n)}-T_A^{(n)}=\frac{|t|^2-|t_A|^2}{|t|^2+|t_A|^2} \; .
\ee
In particular, for $\cos{k_SL}=0$, Eqs.~(\ref{eq:coeffs}, \ref{eq:rescond}, \ref{eq:asympt}) imply that perfect electronic transmission is achieved at integer values of $\nu$, which translates into the conductance peaks shown in Fig.~\ref{FigInterfero}. The blue curve in the inset of Fig.~\ref{FigInterfero} interestingly suggests that these peaks should survive if $\cos{k_SL}$ is not strictly zero.

A closer look at Fig.~\ref{FigInterfero} shows that the conductance peaks are all the more visible that they are accompanied by dips at half-integer values of $\nu$. The simplification brought by the vanishing of the amplitude of crossed Andreev reflection when $\cos{k_SL}=0$ (see Eq.~(\ref{eq:coeffs})) allows to support this observation by an analytical statement, as one can then show that $|h'_N|^2=0$, while $|e'_N|^2 = |t|^2$ if $N$ is odd, and zero otherwise. This yields $T^{(n)}-T_A^{(n)} \leq |t|^2$: in other words, $G$ this time remarkably does not increase with $W$ (red curves in Fig.~\ref{FigRetro}) and only charge carriers having tunneled through the S region at the first vertex may actually contribute to $G$. This even-odd effect is a signature of the destructive interference of electron and hole paths ($\delta\phi=\pi$) when $\nu$ is half-integer, which is already manifest in the limit of a single QH/S interface \cite{Takagaki98,Asano00,Chtchelkatchev01,Hoppe00}. Indeed, taking advantage of the unitarity of the local propagator in this limit, 
its effect on vector $(e_i,h_i)^T$ can then be interpreted as rotating the latter on the Bloch sphere with a frequency $\omega$ given by
$\cos{(\omega/2)}=\cos{\theta_n}\cos{\pi\nu}$, such that, for half-integer values of $\nu$, the rotation frequency is $\omega=\pi$ and the same even-odd effect is displayed.
The red curve in the inset of Fig.~\ref{FigInterfero} however clearly demonstrates that this interpretation breaks down when $\cos{k_SL}$ is no longer zero, as expected from the explicit coupling between $\nu$ and $k_SL$ displayed in Eq.~(\ref{eq:rescond}). In fact, for $\cos{k_SL}=1$, the positions of the conductance peaks and dips are exchanged with respect to the previous situation (see inset of Fig.~\ref{FigInterfero}), with dips occuring at integer values of $\nu$ and peaks at half-integer values (in agreement with Eq.~(\ref{eq:rescond})). The value of the peaks will however be of lesser magnitude in this case, since $t_A\neq0$ (see Eq.~(\ref{eq:asympt})).

The above predictions are in clear contrast with what would be expected from a purely classical point of view in this situation (dashed lines in Figs.~\ref{FigInterfero} and \ref{FigRetro}):
the classical dynamics along the left interface indeed follows a one-dimensional random walk with backward hopping probability $p=|t|^2+|t_A|^2$, for which Eq.~(\ref{eq:GenRec}) can be solved by recurrence, yielding $R^{(n)}+R_A^{(n)} \leq (1-p)/(1-p+Np)$. In other words, transmission through the S region is classically expected to increase -- albeit only slowly (algebraically) -- with $W$, which can be understood as arising from the fact that charge carriers will have to experience an ever larger number of scattering events to cross the system (see Fig.~\ref{FigSetup}). That this is not always the case semiclassically (as discussed above) illustrates the crucial role played by quantum interferences in this setup. 

As a closing remark, let us briefly comment on the sensitivity to disorder of these interference effects. While disorder away from the S region should bring no meaningful change to the results, the presence of strong enough disorder in the vicinity of the QH/S interfaces will essentially randomize the phases of the charge carriers, including the phase difference $\delta\phi$ acquired by electron and hole carriers between consecutive vertices, thus likely spoiling the conductance oscillations depicted in Fig.~\ref{FigInterfero}. However, experimental measurements of the conductance in graphene would still be valuable in the disordered regime, even in the limit of a single QH/S interface, drawing on an analogy between QH/S and bipolar QH junctions which shall be discussed elsewhere \cite{Carmier13NSN}: they could indeed allow estimating the relevance of charge density fluctuations \cite{Martin08} regarding the equipartition of charge carriers observed in bipolar QH junctions \cite{Williams07,Ozyilmaz07,Abanin07}.

To summarize, we have seen that spectacular quantum interference effects can arise at the interface between a QH insulator and a superconductor, and that these effects can be quantitatively understood using an intuitive trajectory-based semiclassical approach. We provided a clear experimental signature of these quantum interferences by showing that the conductance flowing through a QH/S/QH junction should feature characteristic oscillations as a function of the Landau level filling factor $\nu$. We hope our results will motivate further studies of edge state transport at the interface between topologically distinct phases.

\begin{acknowledgments}
\textit{Acknowledgements}.
I am grateful to D. Badiane, G. Fleury, S. Gu\'eron, M. Houzet, T. L\"ofwander, J. Meyer, D. Ullmo and X. Waintal for useful discussions at various stages of this project, and acknowledge financial support from STREP ConceptGraphene.
\end{acknowledgments}

\bibliography{QH-S}

\begin{thebibliography}{34}
\expandafter\ifx\csname natexlab\endcsname\relax\def\natexlab#1{#1}\fi
\expandafter\ifx\csname bibnamefont\endcsname\relax
  \def\bibnamefont#1{#1}\fi
\expandafter\ifx\csname bibfnamefont\endcsname\relax
  \def\bibfnamefont#1{#1}\fi
\expandafter\ifx\csname citenamefont\endcsname\relax
  \def\citenamefont#1{#1}\fi
\expandafter\ifx\csname url\endcsname\relax
  \def\url#1{\texttt{#1}}\fi
\expandafter\ifx\csname urlprefix\endcsname\relax\def\urlprefix{URL }\fi
\providecommand{\bibinfo}[2]{#2}
\providecommand{\eprint}[2][]{\url{#2}}

\bibitem[{\citenamefont{Eroms et~al.}(2005)\citenamefont{Eroms, Weiss, DeBoeck,
  Borghs, and Z\"ulicke}}]{Eroms05}
\bibinfo{author}{\bibfnamefont{J.}~\bibnamefont{Eroms}},
  \bibinfo{author}{\bibfnamefont{D.}~\bibnamefont{Weiss}},
  \bibinfo{author}{\bibfnamefont{J.}~\bibnamefont{DeBoeck}},
  \bibinfo{author}{\bibfnamefont{G.}~\bibnamefont{Borghs}}, \bibnamefont{and}
  \bibinfo{author}{\bibfnamefont{U.}~\bibnamefont{Z\"ulicke}},
  \bibinfo{journal}{Phys. Rev. Lett.} \textbf{\bibinfo{volume}{95}},
  \bibinfo{pages}{107001} (\bibinfo{year}{2005}).

\bibitem[{\citenamefont{Hoppe et~al.}(2000)\citenamefont{Hoppe, Z\"ulicke, and
  Sch\"on}}]{Hoppe00}
\bibinfo{author}{\bibfnamefont{H.}~\bibnamefont{Hoppe}},
  \bibinfo{author}{\bibfnamefont{U.}~\bibnamefont{Z\"ulicke}},
  \bibnamefont{and} \bibinfo{author}{\bibfnamefont{G.}~\bibnamefont{Sch\"on}},
  \bibinfo{journal}{Phys. Rev. Lett.} \textbf{\bibinfo{volume}{84}},
  \bibinfo{pages}{1804} (\bibinfo{year}{2000}).

\bibitem[{\citenamefont{Takayanagi and Akazaki}(1998)}]{Takayanagi98}
\bibinfo{author}{\bibfnamefont{H.}~\bibnamefont{Takayanagi}} \bibnamefont{and}
  \bibinfo{author}{\bibfnamefont{T.}~\bibnamefont{Akazaki}},
  \bibinfo{journal}{Physica} \textbf{\bibinfo{volume}{249}},
  \bibinfo{pages}{462} (\bibinfo{year}{1998}).

\bibitem[{\citenamefont{Moore and Williams}(1999)}]{Moore99}
\bibinfo{author}{\bibfnamefont{T.~D.} \bibnamefont{Moore}} \bibnamefont{and}
  \bibinfo{author}{\bibfnamefont{D.~A.} \bibnamefont{Williams}},
  \bibinfo{journal}{Phys. Rev. B} \textbf{\bibinfo{volume}{59}},
  \bibinfo{pages}{7308} (\bibinfo{year}{1999}).

\bibitem[{\citenamefont{Takagaki}(1998)}]{Takagaki98}
\bibinfo{author}{\bibfnamefont{Y.}~\bibnamefont{Takagaki}},
  \bibinfo{journal}{Phys. Rev. B} \textbf{\bibinfo{volume}{57}},
  \bibinfo{pages}{4009} (\bibinfo{year}{1998}).

\bibitem[{\citenamefont{Asano}(2000)}]{Asano00}
\bibinfo{author}{\bibfnamefont{Y.}~\bibnamefont{Asano}},
  \bibinfo{journal}{Phys. Rev. B} \textbf{\bibinfo{volume}{61}},
  \bibinfo{pages}{1732} (\bibinfo{year}{2000}).

\bibitem[{\citenamefont{Chtchelkatchev}(2001)}]{Chtchelkatchev01}
\bibinfo{author}{\bibfnamefont{N.~M.} \bibnamefont{Chtchelkatchev}},
  \bibinfo{journal}{JETP Lett.} \textbf{\bibinfo{volume}{73}},
  \bibinfo{pages}{94} (\bibinfo{year}{2001}).

\bibitem[{\citenamefont{Novoselov et~al.}(2005)\citenamefont{Novoselov, Geim,
  Morozov, Jiang, Katsnelson, Grigorieva, Dubonos, and Firsov}}]{Novoselov05}
\bibinfo{author}{\bibfnamefont{K.~S.} \bibnamefont{Novoselov}},
  \bibinfo{author}{\bibfnamefont{A.~K.} \bibnamefont{Geim}},
  \bibinfo{author}{\bibfnamefont{S.~V.} \bibnamefont{Morozov}},
  \bibinfo{author}{\bibfnamefont{D.}~\bibnamefont{Jiang}},
  \bibinfo{author}{\bibfnamefont{M.~I.} \bibnamefont{Katsnelson}},
  \bibinfo{author}{\bibfnamefont{I.~V.} \bibnamefont{Grigorieva}},
  \bibinfo{author}{\bibfnamefont{S.~V.} \bibnamefont{Dubonos}},
  \bibnamefont{and} \bibinfo{author}{\bibfnamefont{A.~A.}
  \bibnamefont{Firsov}}, \bibinfo{journal}{Nature}
  \textbf{\bibinfo{volume}{438}}, \bibinfo{pages}{197} (\bibinfo{year}{2005}).

\bibitem[{\citenamefont{Zhang et~al.}(2005)\citenamefont{Zhang, Tan, Stormer,
  and Kim}}]{Zhang05}
\bibinfo{author}{\bibfnamefont{Y.}~\bibnamefont{Zhang}},
  \bibinfo{author}{\bibfnamefont{Y.-W.} \bibnamefont{Tan}},
  \bibinfo{author}{\bibfnamefont{H.~L.} \bibnamefont{Stormer}},
  \bibnamefont{and} \bibinfo{author}{\bibfnamefont{P.}~\bibnamefont{Kim}},
  \bibinfo{journal}{Nature} \textbf{\bibinfo{volume}{438}},
  \bibinfo{pages}{201} (\bibinfo{year}{2005}).

\bibitem[{\citenamefont{Heersche et~al.}(2007)\citenamefont{Heersche,
  Jarillo-Herrero, Oostinga, Vandersypen, and Morpurgo}}]{Heersche07}
\bibinfo{author}{\bibfnamefont{H.~B.} \bibnamefont{Heersche}},
  \bibinfo{author}{\bibfnamefont{P.}~\bibnamefont{Jarillo-Herrero}},
  \bibinfo{author}{\bibfnamefont{J.~B.} \bibnamefont{Oostinga}},
  \bibinfo{author}{\bibfnamefont{L.~M.~K.} \bibnamefont{Vandersypen}},
  \bibnamefont{and} \bibinfo{author}{\bibfnamefont{A.~F.}
  \bibnamefont{Morpurgo}}, \bibinfo{journal}{Nature}
  \textbf{\bibinfo{volume}{446}}, \bibinfo{pages}{56} (\bibinfo{year}{2007}).

\bibitem[{\citenamefont{Komatsu et~al.}(2012)\citenamefont{Komatsu, Li,
  Autier-Laurent, Bouchiat, and Gu\'eron}}]{Komatsu12}
\bibinfo{author}{\bibfnamefont{K.}~\bibnamefont{Komatsu}},
  \bibinfo{author}{\bibfnamefont{C.}~\bibnamefont{Li}},
  \bibinfo{author}{\bibfnamefont{S.}~\bibnamefont{Autier-Laurent}},
  \bibinfo{author}{\bibfnamefont{H.}~\bibnamefont{Bouchiat}}, \bibnamefont{and}
  \bibinfo{author}{\bibfnamefont{S.}~\bibnamefont{Gu\'eron}},
  \bibinfo{journal}{Phys. Rev. B} \textbf{\bibinfo{volume}{86}},
  \bibinfo{pages}{115412} (\bibinfo{year}{2012}).

\bibitem[{\citenamefont{Popinciuc et~al.}(2012)\citenamefont{Popinciuc, Calado,
  Liu, Akhmerov, Klapwijk, and Vandersypen}}]{Popinciuc12}
\bibinfo{author}{\bibfnamefont{M.}~\bibnamefont{Popinciuc}},
  \bibinfo{author}{\bibfnamefont{V.~E.} \bibnamefont{Calado}},
  \bibinfo{author}{\bibfnamefont{X.~L.} \bibnamefont{Liu}},
  \bibinfo{author}{\bibfnamefont{A.~R.} \bibnamefont{Akhmerov}},
  \bibinfo{author}{\bibfnamefont{T.~M.} \bibnamefont{Klapwijk}},
  \bibnamefont{and} \bibinfo{author}{\bibfnamefont{L.~M.~K.}
  \bibnamefont{Vandersypen}}, \bibinfo{journal}{Phys. Rev. B}
  \textbf{\bibinfo{volume}{85}}, \bibinfo{pages}{205404}
  (\bibinfo{year}{2012}).

\bibitem[{\citenamefont{Rickhaus et~al.}(2012)\citenamefont{Rickhaus, Weiss,
  Marot, and Sch\"onenberger}}]{Rickhaus12}
\bibinfo{author}{\bibfnamefont{P.}~\bibnamefont{Rickhaus}},
  \bibinfo{author}{\bibfnamefont{M.}~\bibnamefont{Weiss}},
  \bibinfo{author}{\bibfnamefont{L.}~\bibnamefont{Marot}}, \bibnamefont{and}
  \bibinfo{author}{\bibfnamefont{C.}~\bibnamefont{Sch\"onenberger}},
  \bibinfo{journal}{Nano Letters} \textbf{\bibinfo{volume}{12}},
  \bibinfo{pages}{1942} (\bibinfo{year}{2012}).

\bibitem[{\citenamefont{Hasan and Kane}(2010)}]{Hasan10}
\bibinfo{author}{\bibfnamefont{M.~Z.} \bibnamefont{Hasan}} \bibnamefont{and}
  \bibinfo{author}{\bibfnamefont{C.~L.} \bibnamefont{Kane}},
  \bibinfo{journal}{Rev. Mod. Phys.} \textbf{\bibinfo{volume}{82}},
  \bibinfo{pages}{3045} (\bibinfo{year}{2010}).

\bibitem[{\citenamefont{Qi and Zhang}(2011)}]{Qi11}
\bibinfo{author}{\bibfnamefont{X.-L.} \bibnamefont{Qi}} \bibnamefont{and}
  \bibinfo{author}{\bibfnamefont{S.-C.} \bibnamefont{Zhang}},
  \bibinfo{journal}{Rev. Mod. Phys.} \textbf{\bibinfo{volume}{83}},
  \bibinfo{pages}{1057} (\bibinfo{year}{2011}).

\bibitem[{\citenamefont{Chtchelkatchev and
  Burmistrov}(2007)}]{Chtchelkatchev07}
\bibinfo{author}{\bibfnamefont{N.~M.} \bibnamefont{Chtchelkatchev}}
  \bibnamefont{and} \bibinfo{author}{\bibfnamefont{I.~S.}
  \bibnamefont{Burmistrov}}, \bibinfo{journal}{Phys. Rev. B}
  \textbf{\bibinfo{volume}{75}}, \bibinfo{pages}{214510}
  (\bibinfo{year}{2007}).

\bibitem[{\citenamefont{Carmier et~al.}(2010)\citenamefont{Carmier, Lewenkopf,
  and Ullmo}}]{Carmier10}
\bibinfo{author}{\bibfnamefont{P.}~\bibnamefont{Carmier}},
  \bibinfo{author}{\bibfnamefont{C.}~\bibnamefont{Lewenkopf}},
  \bibnamefont{and} \bibinfo{author}{\bibfnamefont{D.}~\bibnamefont{Ullmo}},
  \bibinfo{journal}{Phys. Rev. B (R)} \textbf{\bibinfo{volume}{81}},
  \bibinfo{pages}{241406} (\bibinfo{year}{2010}).

\bibitem[{\citenamefont{Carmier et~al.}(2011)\citenamefont{Carmier, Lewenkopf,
  and Ullmo}}]{Carmier11}
\bibinfo{author}{\bibfnamefont{P.}~\bibnamefont{Carmier}},
  \bibinfo{author}{\bibfnamefont{C.}~\bibnamefont{Lewenkopf}},
  \bibnamefont{and} \bibinfo{author}{\bibfnamefont{D.}~\bibnamefont{Ullmo}},
  \bibinfo{journal}{Phys. Rev. B} \textbf{\bibinfo{volume}{84}},
  \bibinfo{pages}{195428} (\bibinfo{year}{2011}).

\bibitem[{\citenamefont{Rickhaus et~al.}(2013)\citenamefont{Rickhaus, Maurand,
  Liu, Weiss, Richter, and Sch\"onenberger}}]{Rickhaus13}
\bibinfo{author}{\bibfnamefont{P.}~\bibnamefont{Rickhaus}},
  \bibinfo{author}{\bibfnamefont{R.}~\bibnamefont{Maurand}},
  \bibinfo{author}{\bibfnamefont{M.~H.} \bibnamefont{Liu}},
  \bibinfo{author}{\bibfnamefont{M.}~\bibnamefont{Weiss}},
  \bibinfo{author}{\bibfnamefont{K.}~\bibnamefont{Richter}}, \bibnamefont{and}
  \bibinfo{author}{\bibfnamefont{C.}~\bibnamefont{Sch\"onenberger}},
  \bibinfo{journal}{arXiv:1304.6590}  (\bibinfo{year}{2013}).

\bibitem[{\citenamefont{Grushina et~al.}(2013)\citenamefont{Grushina, Ki, and
  Morpurgo}}]{Grushina13}
\bibinfo{author}{\bibfnamefont{A.~L.} \bibnamefont{Grushina}},
  \bibinfo{author}{\bibfnamefont{D.-K.} \bibnamefont{Ki}}, \bibnamefont{and}
  \bibinfo{author}{\bibfnamefont{A.~F.} \bibnamefont{Morpurgo}},
  \bibinfo{journal}{Appl. Phys. Lett.} \textbf{\bibinfo{volume}{102}},
  \bibinfo{pages}{223102} (\bibinfo{year}{2013}).

\bibitem[{\citenamefont{Mizuno et~al.}(2013)\citenamefont{Mizuno, Nielsen, and
  Du}}]{Mizuno13}
\bibinfo{author}{\bibfnamefont{N.}~\bibnamefont{Mizuno}},
  \bibinfo{author}{\bibfnamefont{B.}~\bibnamefont{Nielsen}}, \bibnamefont{and}
  \bibinfo{author}{\bibfnamefont{X.}~\bibnamefont{Du}},
  \bibinfo{journal}{arXiv:1305.2180}  (\bibinfo{year}{2013}).

\bibitem[{\citenamefont{B\"uttiker et~al.}(1985)\citenamefont{B\"uttiker, Imry,
  Landauer, and Pinhas}}]{Buettiker85}
\bibinfo{author}{\bibfnamefont{M.}~\bibnamefont{B\"uttiker}},
  \bibinfo{author}{\bibfnamefont{Y.}~\bibnamefont{Imry}},
  \bibinfo{author}{\bibfnamefont{R.}~\bibnamefont{Landauer}}, \bibnamefont{and}
  \bibinfo{author}{\bibfnamefont{S.}~\bibnamefont{Pinhas}},
  \bibinfo{journal}{Phys. Rev. B} \textbf{\bibinfo{volume}{31}},
  \bibinfo{pages}{6207} (\bibinfo{year}{1985}).

\bibitem[{\citenamefont{Blonder et~al.}(1982)\citenamefont{Blonder, Tinkham,
  and Klapwijk}}]{Blonder82}
\bibinfo{author}{\bibfnamefont{G.~E.} \bibnamefont{Blonder}},
  \bibinfo{author}{\bibfnamefont{M.}~\bibnamefont{Tinkham}}, \bibnamefont{and}
  \bibinfo{author}{\bibfnamefont{T.~M.} \bibnamefont{Klapwijk}},
  \bibinfo{journal}{Phys. Rev. B} \textbf{\bibinfo{volume}{25}},
  \bibinfo{pages}{4515} (\bibinfo{year}{1982}).

\bibitem[{\citenamefont{Rakyta et~al.}(2010)\citenamefont{Rakyta, Kormanyos,
  Cserti, and Koskinen}}]{Rakyta10}
\bibinfo{author}{\bibfnamefont{P.}~\bibnamefont{Rakyta}},
  \bibinfo{author}{\bibfnamefont{A.}~\bibnamefont{Kormanyos}},
  \bibinfo{author}{\bibfnamefont{J.}~\bibnamefont{Cserti}}, \bibnamefont{and}
  \bibinfo{author}{\bibfnamefont{P.}~\bibnamefont{Koskinen}},
  \bibinfo{journal}{Phys. Rev. B} \textbf{\bibinfo{volume}{81}},
  \bibinfo{pages}{115411} (\bibinfo{year}{2010}).

\bibitem[{\citenamefont{Akhmerov and Beenakker}(2007)}]{Akhmerov07}
\bibinfo{author}{\bibfnamefont{A.~R.} \bibnamefont{Akhmerov}} \bibnamefont{and}
  \bibinfo{author}{\bibfnamefont{C.~W.~J.} \bibnamefont{Beenakker}},
  \bibinfo{journal}{Phys. Rev. Lett.} \textbf{\bibinfo{volume}{98}},
  \bibinfo{pages}{157003} (\bibinfo{year}{2007}).

\bibitem[{\citenamefont{Carmier}(2013)}]{Carmier13NSN}
\bibinfo{author}{\bibfnamefont{P.}~\bibnamefont{Carmier}}, \bibinfo{journal}{in
  preparation}  (\bibinfo{year}{2013}).

\bibitem[{\citenamefont{Fisher and Lee}(1981)}]{Fisher81}
\bibinfo{author}{\bibfnamefont{D.~S.} \bibnamefont{Fisher}} \bibnamefont{and}
  \bibinfo{author}{\bibfnamefont{P.~A.} \bibnamefont{Lee}},
  \bibinfo{journal}{Phys. Rev. B} \textbf{\bibinfo{volume}{23}},
  \bibinfo{pages}{6851} (\bibinfo{year}{1981}).

\bibitem[{\citenamefont{Baranger and Stone}(1989)}]{Baranger89}
\bibinfo{author}{\bibfnamefont{H.~U.} \bibnamefont{Baranger}} \bibnamefont{and}
  \bibinfo{author}{\bibfnamefont{A.~D.} \bibnamefont{Stone}},
  \bibinfo{journal}{Phys. Rev. B} \textbf{\bibinfo{volume}{40}},
  \bibinfo{pages}{8169} (\bibinfo{year}{1989}).

\bibitem[{\citenamefont{Beenakker}(2006)}]{Beenakker06}
\bibinfo{author}{\bibfnamefont{C.~W.~J.} \bibnamefont{Beenakker}},
  \bibinfo{journal}{Phys. Rev. Lett.} \textbf{\bibinfo{volume}{97}},
  \bibinfo{pages}{067007} (\bibinfo{year}{2006}).

\bibitem[{\citenamefont{Tinkham}(2004)}]{Tinkham}
\bibinfo{author}{\bibfnamefont{M.}~\bibnamefont{Tinkham}},
  \emph{\bibinfo{title}{Introduction to Superconductivity}}
  (\bibinfo{publisher}{Dover}, \bibinfo{year}{2004}).

\bibitem[{\citenamefont{Martin et~al.}(2008)\citenamefont{Martin, Akerman,
  Ulbricht, Lohmann, Smet, von Klitzing, and Yacoby}}]{Martin08}
\bibinfo{author}{\bibfnamefont{J.}~\bibnamefont{Martin}},
  \bibinfo{author}{\bibfnamefont{N.}~\bibnamefont{Akerman}},
  \bibinfo{author}{\bibfnamefont{G.}~\bibnamefont{Ulbricht}},
  \bibinfo{author}{\bibfnamefont{T.}~\bibnamefont{Lohmann}},
  \bibinfo{author}{\bibfnamefont{J.~H.} \bibnamefont{Smet}},
  \bibinfo{author}{\bibfnamefont{K.}~\bibnamefont{von Klitzing}},
  \bibnamefont{and} \bibinfo{author}{\bibfnamefont{A.}~\bibnamefont{Yacoby}},
  \bibinfo{journal}{Nat. Phys.} \textbf{\bibinfo{volume}{4}},
  \bibinfo{pages}{144} (\bibinfo{year}{2008}).

\bibitem[{\citenamefont{Williams et~al.}(2007)\citenamefont{Williams, DiCarlo,
  and Marcus}}]{Williams07}
\bibinfo{author}{\bibfnamefont{J.~R.} \bibnamefont{Williams}},
  \bibinfo{author}{\bibfnamefont{L.}~\bibnamefont{DiCarlo}}, \bibnamefont{and}
  \bibinfo{author}{\bibfnamefont{C.~M.} \bibnamefont{Marcus}},
  \bibinfo{journal}{Science} \textbf{\bibinfo{volume}{317}},
  \bibinfo{pages}{638} (\bibinfo{year}{2007}).

\bibitem[{\citenamefont{Ozyilmaz et~al.}(2007)\citenamefont{Ozyilmaz,
  Jarillo-Herrero, Efetov, Abanin, Levitov, and Kim}}]{Ozyilmaz07}
\bibinfo{author}{\bibfnamefont{B.}~\bibnamefont{Ozyilmaz}},
  \bibinfo{author}{\bibfnamefont{P.}~\bibnamefont{Jarillo-Herrero}},
  \bibinfo{author}{\bibfnamefont{D.}~\bibnamefont{Efetov}},
  \bibinfo{author}{\bibfnamefont{D.~A.} \bibnamefont{Abanin}},
  \bibinfo{author}{\bibfnamefont{L.~S.} \bibnamefont{Levitov}},
  \bibnamefont{and} \bibinfo{author}{\bibfnamefont{P.}~\bibnamefont{Kim}},
  \bibinfo{journal}{Phys.~Rev.~Lett.} \textbf{\bibinfo{volume}{99}},
  \bibinfo{pages}{166804} (\bibinfo{year}{2007}).

\bibitem[{\citenamefont{Abanin and Levitov}(2007)}]{Abanin07}
\bibinfo{author}{\bibfnamefont{D.~A.} \bibnamefont{Abanin}} \bibnamefont{and}
  \bibinfo{author}{\bibfnamefont{L.~S.} \bibnamefont{Levitov}},
  \bibinfo{journal}{Science} \textbf{\bibinfo{volume}{317}},
  \bibinfo{pages}{641} (\bibinfo{year}{2007}).

\end{thebibliography}
\bibliographystyle{apsrev}

\end{document}